\documentclass[a4paper,fleqn,usenatbib]{mnras}

\usepackage{mathptmx}

\usepackage[T1]{fontenc}
\usepackage{ae,aecompl}

\usepackage{graphicx}   
\usepackage{amsmath}    
\usepackage{amssymb}    

\usepackage{booktabs}

\usepackage{multirow}

               {           \\\bottomrule%
                     \end{tabular}%
                   \end{small}
               \end{center} 
               }

\usepackage{dblfloatfix}

\usepackage[usenames]{color}

\usepackage[normalem]{ulem}

\newcommand{\be}{\begin{equation}}
\newcommand{\ee}{\end{equation}}
\newcommand{\bi}{\begin{itemize}}
\newcommand{\ei}{\end{itemize}}

\title[The shrinking domain framework]{The shrinking domain framework I:  a new, faster, more efficient approach to cosmological simulations}

\author[Llinares, C.]{
Claudio Llinares,$^{1}$\thanks{E-mail: claudio.llinares@durham.ac.uk}\\
$^{1}$Institute for Computational Cosmology, Department of Physics, Durham University, Durham DH1 3LE, U.K.
}

\date{Accepted XXX. Received YYY; in original form ZZZ}

\pubyear{2017}

\begin{document}
\label{firstpage}
\pagerange{\pageref{firstpage}--\pageref{lastpage}}
\maketitle

\begin{abstract}
The advent of the new generation of wide field galaxy surveys makes large N-body cosmological simulations a necessary evil.  While the cosmological simulation codes have evolved a lot since the first calculations in the 80s, the computational requirements for generating data that is relevant for large surveys remain challenging.  We propose an alternative approach that can speed up these simulations.  The framework is based on the idea of reducing the size of the integration region following the lightcone of an observer at redshift zero and thus simulating only the parts of the Universe that can be observed.  A possible implementation of this framework is presented, as well as tests of its accuracy and performance.  These simple tests, based on conservative assumptions, show that the new framework gives a factor of three speed up with respect to the usual approach.
\end{abstract}




\section{Introduction}

Cosmological simulations were first used in the early 1980's to satisfy the need to make predictions of the cosmological evolution in the non-linear regime, where the growth equation can not be linearised with respect to the density fluctuations.  The codes have greatly evolved since these early times \citep{1998ARA&A..36..599B,2012PDU.....1...50K}, including refined tree algorithms for calculation of forces, adaptive mesh refinements, MPI paralelization, baryonic physics, etc.  Owing to this complexity, the number of code lines in a typical N-body code grew from a few hundred to more than a hundred thousand.  However, there is one thing that was not affected by these developments:  the set of equations that all these codes solve today is exactly the same as in the original cases.  These are Hamilton equations for the position of the N-body particles plus Einstein's equations to first order in the metric perturbations and in the quasi-static limit (i.e. Poisson's equation).  The boundary conditions are periodic in a cubic box of constant comoving size.

While the method is known to be accurate \citep{2016JCAP...04..047S}, there are several reasons why we may want to explore alternative approaches.  Before getting into them, I describe what the new framework proposed here consists of.  The question we need to ask to define this new framework is ``What is the smallest box required for a simulation to be considered sufficiently accurate for a given scientific question?''.  The answer today is based on a compromise between the physics that we want to simulate and the available computational resources.  The larger the box, the larger the modes we will be able to simulate, but the lower the resolution for a fixed number of particles.  Increasing the number of particles will come with a computational cost for which resources are not always available.  This problem can be alleviated if we take into account causality.  For instance, if we want to make predictions for a given galaxy survey, we will need to simulate a box that contains all the modes that exist in the survey.  However, the finite speed of propagation of information tells us that we only need to simulate the regions that are causality connected with the galaxies that exist in the survey.  This means that the comoving size of the simulation box does not need to be fixed in time, but can be reduced as time passes and the lightcone of an observer at redshift zero shrinks.

\begin{figure*}
  \begin{center}
    \includegraphics[width=\textwidth]{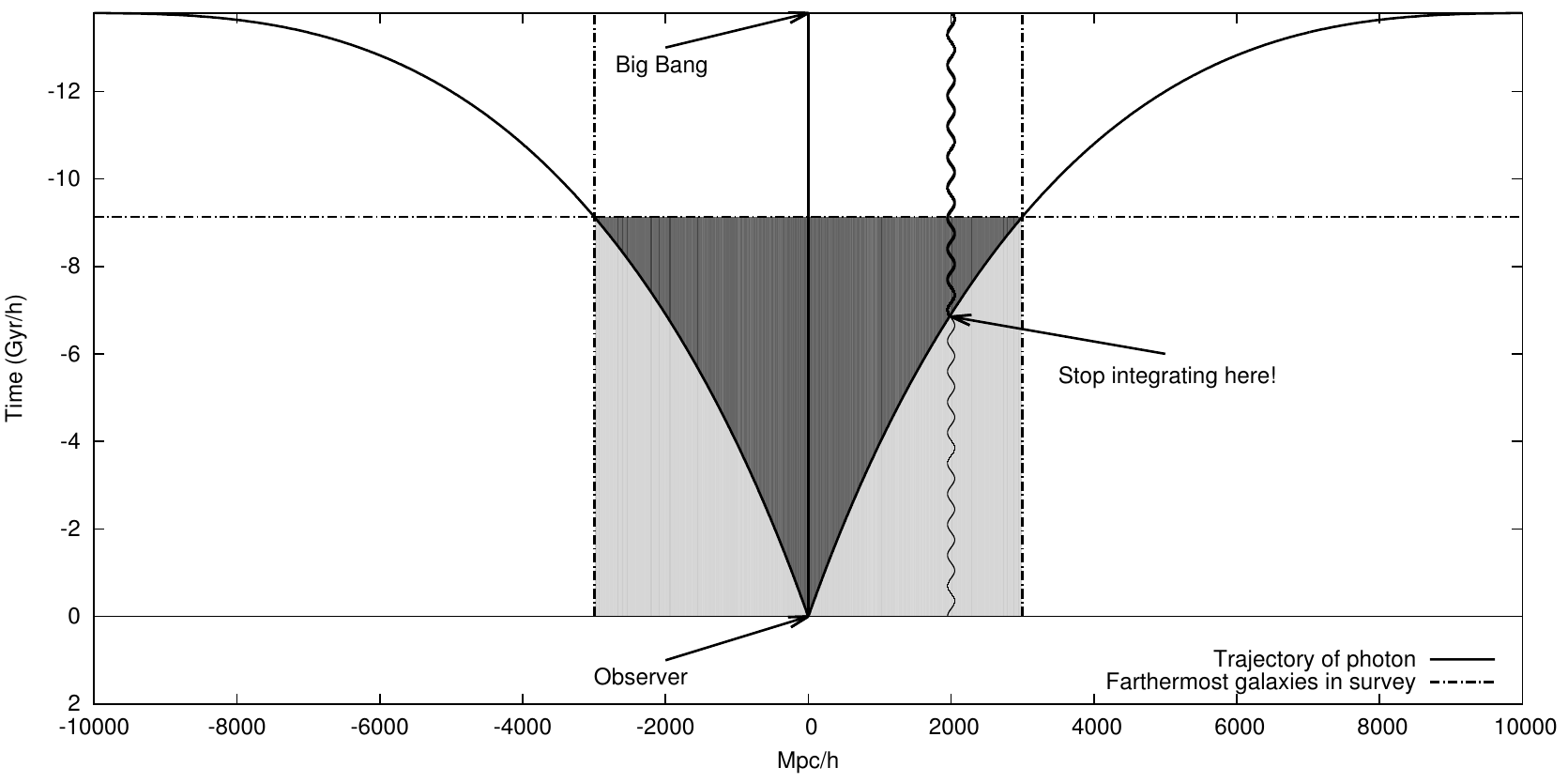}
    \caption{Space-time diagram of a universe with one spatial dimension.  See text for explanation.} 
    \label{fig:explaining_method}
  \end{center}
\end{figure*}

Figure \ref{fig:explaining_method} shows a concrete example in a space-time diagram of a universe with one spatial dimension.  The vertical and horizontal axes correspond to comoving coordinates in space and time.  Newtonian time runs backwards towards the upper direction.  The continuous line shows the past lightcone of an observer at redshift zero, which can be calculated as the trajectory of a two photons that arrive at an observer today.  The horizontal line shows the position in time of the farthest galaxies included in a hypothetical survey.  The vertical lines correspond to the limits of the simulation box that should be used if we want to contain the whole survey in the box.  Finally, the wavy line is the world line of a typical galaxy in the survey, which in this case is located at 2000 Mpc/h from the observer at redshift zero.  The galaxy moves out of the observer's lightcone at $t\sim 6$ Gyr$/h$.  Following its dynamics after this moment is not necessary from an observer's perspective because it will not be observable.  Furthermore, and more importantly, the object becomes causally disconected from everything inside the lightcone and thus its existence will have no impact on anything that can be observed today.  A different way of saying the same thing is the following:  the survey will observe only galaxies lying in the dark shaded region in Fig \ref{fig:explaining_method}.  However, the current simulation technique follows the dynamics of all the galaxies that lie within the dark and light shaded regions together.  The use of resources can be optimized by simulating only what happens in the dark region of the space-time diagram.  The newly released resources can be used for instance to simulate larger scales at higher redshift (which are still inside the lightcone and thus affect the galaxies within the survey) or for increasing the resolution in the observable region.

\begin{table*} 
  \centering
  \caption{Differences between the usual algorithm and the new one proposed here.  Particle mesh codes are adopted as being representative of the usual algorithm.  Differences with tree codes are related to details of the force calculation, but not to the basic definition of the standard N-body method.}
  \label{table:differences_methods}
\begin{tabular}{ | r | p{0.3\textwidth} | p{0.3\textwidth} |}
\toprule
  &  Usual algorithm  & Shrinking domain framework  (i.e. this paper) \\
\midrule \midrule
    Boundary conditions & Periodic. & Open. \\
\midrule
    Box size & Cubical.  Size is fixed after factoring out the expansion. & Spherical.  Size is variable following the past light cone of an observer at redshift zero. \\
\midrule
    Forces and gravitational potential  & Forces obtained by direct summation or as a discretization of the derivatives of the potential, which is obtained from Poisson's equation (in the quasi-static limit). & Fields (potential or forces) are evolved in time following solution of linearised Einstein's equations.   Several options exist to write the equations. \\
\midrule
Initial conditions for the gravity solver & Not needed.  & Initial conditions for the fields that are evolved in time can be obtained from linear theory.  \\
\midrule
Natural outcomes of the gravity calculations & Gravitational potential and its gradient & Gravitational potential, its gradients and its time derivative. \\
\midrule
Output format & Lightcones or snapshots of fixed comoving size. & Light cone or snapshots of different comoving sizes. \\
\bottomrule
\end{tabular}
\end{table*}

So the method consists of simulating \textit{only} what happens inside the past lightcone of an observer at redshift zero.  To further clarify this new framework, I summarize a few of its characteristics:
\bi
\item The box is not a cube, but a sphere which reduces in radius as time passes.  Note that this respects the symmetry of the observable universe.
\item As the volume is a sphere and shrinks with time, the boundary conditions can not be periodic.  We are now forced to work with open boundary conditions.  This again is more consistent with what happens in the real universe and thus, more realistic.
\item The gravity calculation must take into account that objects outside the lightcone do not affect objects that can be observed at redshift zero.  To do this, we need to drop the quasi-static approximation (i.e. we need to take into account that information travels at a finite speed as in the real universe).  Thus, we need a potential solver that allows spherical volume and open boundary conditions.  We will get physically well motivated equations that can do this by moving one order up in a perturbative expansion of Einstein's equations, which include time derivatives of a scalar perturbation of the metric.
\item Within the new framework, it is not longer possible to write snapshots of the complete volume at fixed redshifts.  The natural output under the new framework is a lightcone.  This may seem to be a restriction of the method, however, it will force us to do analyses of the simulations that are closer to what observers do and thus will make our predictions more realistic.  Note that it will still be possible to store volumes with different sizes at fixed redshifts.
\item The method provides a faster way of running simulations (subsequent sections will quantify how fast the simulations become), but it is not approximate.  The complete implementation of the method will actually be more accurate than the standard approach because it relaxes the quasi-static approximation and will not be affected by the periodicity of the box.
\ei
Table \ref{table:differences_methods} summarizes the main differences between the standard algorithm and the shrinking domain framework proposed here.

Coming back to the question of why we may want to reinvent the wheel, here are some advantages of the new framework:
\bi
\item The range of scales simulated today is limited by technicalities of the simulation and not by the physics.  The method proposed here permits us to focus the computational resources to simulate the minimum information necessary, so the rest of the resources can be dedicated to increase resolution.  In other words, the new method provides the possibility of obtaining a better light cone with the same computational resources.
\item While the technique seems to be restricted only to surveys, where a lightcone of particles are required, it is also useful for simulations that have a theoretical purpose.  Theoreticians do not need a lightcone, but a reasonable sample of galaxies to do statistics.  The problem is that today's simulations are so large that the samples become impractical, accounting for several tera-bytes per snapshot.  We need to find a way to make the samples smaller using well informed criteria to get rid of galaxies.  The criteria proposed here are based on the observations:  we will not simulate galaxies that can not be observed.  In this case, the error bars on the samples will be more realistic because they are based on observable samples.  
\item As the simulation volume is given by the physics, the result of the simulations is less dependent on the parameters of the simulation (in particular the box size and resolution).  These simulations will always have the largest box size they require (and no more than that).
\item Everything we learned from cosmological simulations was calculated with the standard method.  Results obtained with an alternative approach can be used to make the ultimate test of the codes.
\ei

The rest of the paper describes a particular implementation of the new framework (Section \ref{section:implementing}), the tests that it passes and measurements of its performance compared to that of the standard algorithm (Section \ref{section:tests}).  The conclusions are presented in Section \ref{section:conclusions}.

\section{Implementing the shrinking domain framework}
\label{section:implementing}

The reason why I talk about a new \textit{framework} instead of a new \textit{method} is that the specific method used to solve the N-body equations (or even the equations themselves) is not fixed.  Only the geometry of the simulated volume is.  So in order to make a complete implementation of the framework, we need to fix both the N-body equations and the method that is used to solve them.  Any set of equations that allow us to work with a spherical domain that shrinks with time will do the job.  I describe here one possibility.

\subsection{Equations for the N-body particles and gravity}

The implementation presented here is based on the particle mesh technique, which means that the N-body code will track the positions and velocities of particles and use an auxiliary grid to calculate the gravitational potential and forces.

The equations solved by standard particle mesh codes are Hamilton equations for the position and velocity of each particle and Poisson's equation for the gravitational potential:
\begin{align}
\label{geo_1}
\dot{\textbf{x}}_i & = \frac{\textbf{p}_i}{a} \\
\label{geo_2}
\dot{\textbf{p}}_i & = a \nabla\Phi\\
\nabla^2\Phi & = \frac{3H_0^2\Omega_m}{2a}\delta
\end{align}
where $\textbf{x}_i$ and $\textbf{p}_i$ are the comoving position and momentum of the particle $i$, $a$ is the expansion factor, the dots are derivatives with respect to conformal time, the symbol $\nabla$ is a 3D gradient in comoving coordinates, $\Phi$ is a scalar perturbation of the metric (i.e. the gravitational potential), $H_0$ is the Hubble parameter at redshift zero, $\Omega_m$ is the mean density of the Universe in terms of the critical density and $\delta$ is the perturbation in the density in terms of the mean density of the Universe.

We need to re-write these equations in a way that they can be applied to a system that has open boundary conditions (instead of periodic) and that takes into account the finite speed of propagation of information.  This can be done simply by replacing Poisson's equation by a diffusion equation: 
\be
\label{diffusion}
3 H \dot{\Phi} - c^2 \nabla^2\Phi = -c^2\frac{3H_0^2\Omega_m}{2a}\delta, 
\ee
where $H=\dot{a}/a$.  This new equation can be derived in a self-consistent way starting from Einstein's equation, which I show in Appendix \ref{appendix:equations}.

\subsection{Evolving the particle positions}

The method of choice to solve Eqs. \ref{geo_1} and \ref{geo_2} to obtain the trajectory of the particles is a standard leap-frog scheme \citep{1988csup.book.....H}.  The method consists in advancing positions and velocities using a second order formula.  Conservation of energy can be ensured by displacing the positions and velocities by half a time step from each other:
\begin{align}
\textbf{x}_i(t+dt/2) & = \textbf{x}_i(t-dt/2) + \frac{\textbf{p}_i(t)}{a(t)} dt \\
\textbf{p}_i(t+dt) & = \textbf{p}_i(t) + a(t+dt/2)(\nabla\Phi)_{t+dt/2} dt
\end{align}
Note that this scheme does not allow for variable time steps, which are more efficient in cosmology.  So I use a common modification of the leap-frog \citep{1997astro.ph.10043Q}, which synchronises positions and velocities at the beginning of every time step and advances both quantities using different steps:
\begin{align}
\textbf{x}_i(t+dt/2) & = \textbf{x}_i(t) + \frac{\textbf{p}_i(t)}{a(t)} dt/2\\
\textbf{p}_i(t+dt) & = \textbf{p}_i(t) + a(t+dt/2)(\nabla\Phi)_{t+dt/2} dt \\
\textbf{x}_i(t+dt) & = \textbf{x}_i(t+dt/2) + \frac{\textbf{p}_i(t+dt)}{a(t+dt)} dt/2.
\end{align}
For this to work, we need to obtain the value of the potential at time $t+dt/2$, which we do by evolving the solution of Eq. \ref{diffusion} on the grid and interpolating back to the position of the particles using a CIC scheme.

\subsection{Solving the diffusion equation for gravity}

The solutions for gravity will be obtained with an alternate-direction-explicit (ADE) method \citep[e.g.][]{ade_wuppertal, Aono:2010:NFA:1899721.1899750}.  The method consists of discretizing the equations in space and time and evolving the solution by writing the potential in $t+dt$ at a given cell $(i,j,k)$ as a function of the old value and its neighbours.  Other explicit methods such as the leap-frog scheme can give accurate solutions but have the limitation that are conditional stable, which means that it can be shown that they are stable given a Courant-Friedrichs-Lewy (CFL) condition on the time step.  When applied to the case of gravity, these methods require an unreasonably large number of time steps \citep[e.g.][]{2013PhRvL.110p1101L,2014PhRvD..89h4023L, 2016JCAP...07..053A}.  The ADE method overcomes this problem by making the iterations in two consecutive steps with alternate directions, which makes the method unconditionally stable \citep[e.g.][]{ade_wuppertal}.  So the approach has the main advantage of explicit methods (i.e. simplicity), but at the same time is more stable that other explicit techniques.

To be able to synchronize the potential with the particle's position and velocity at the beginning of the time step, and thus, be allowed to implement variable time steps, the code evolves the potential in two sub-steps of half a time step each.  This means that the values of the potential are updated at the same time the positions are updated.  The ordering of operations in a given time step is summarized in Table \ref{table:method}.

\begin{table*} 
  \centering
  \caption{Comparison of the ordering of operations over a simple step in the usual algorithm and the implementation presented here of the shrinking domain framework.  The non-static algorithm seems more complex, but it does not need to solve Poisson's equation, which is the expensive bit in standard cosmological codes.}
  \label{table:method}
\begin{tabular}{c | c}
\toprule
  Usual algorithm & Shrinking domain (this particular implementation) \\
\midrule \midrule
Advance $\textbf{x}_i$ from $t$ to $t+dt/2$ & Advance $\textbf{x}_i$ from $t$ to $t+dt/2$\\
$\Downarrow$  & $\Downarrow$ \\
Calculate $\delta$, $\Phi$ and $\nabla\Phi$ at $t+dt/2$ & Calculate $\delta$ at $t+dt/2$ \\
$\|$ & $\Downarrow$ \\
$\|$ & Advance $\Phi$ from $t$ to $t+dt/2$\\
$\|$ & $\Downarrow$ \\
$\|$ & Calculate $\nabla \Phi$ from $\Phi$ at $t+dt/2$\\
$\Downarrow$ & $\Downarrow$ \\
Advance $\textbf{p}_i$ from $t$ to $t+dt$ & Advance $\textbf{p}_i$ from $t$ to $t+dt$\\
$\Downarrow$ & $\Downarrow$ \\
Advance $\textbf{x}_i$ from $t+dt/2$ to $t+dt$ & Advance $\textbf{x}_i$ from $t+dt/2$ to $t+dt$\\
$\|$ & $\Downarrow$ \\
$\|$ & Advance $\Phi$ from $t+dt/2$ to $t+dt$\\
$\Downarrow$ & $\Downarrow$ \\
Advance time  from $t$ to $t+dt$ & Advance time  from $t$ to $t+dt$ \\
. & $\Downarrow$ \\
. & Calculate position of lightcone and eliminate particles and cells\\
\bottomrule
\end{tabular}
\end{table*}

\subsection{The code and technical details of the implementation of the new method}

The new framework proposed here may be appealing from a theoretical perspective, but it is useless if not tested.  The code that I use to this end is called \texttt{Solve} \citep{llinares_thesis} and was written with the aim of having a test bench for development of new algorithms.  The code includes static solvers for several theories of gravity, which are Fourier based for the GR case and multigrid for modified gravity (which is required because the equations involved are typically non-linear).   It also includes a non-static solver for modified gravity based on a leap-frog method \citep{2013PhRvL.110p1101L}.  Furthermore, the package includes routines for generating Gaussian random fields, initial conditions for cosmological simulations as well as a power spectrum calculator, which corrects for discreteness effects using the algorithm presented in \cite{2005ApJ...620..559J}.

The original gravity solvers were tested against analytic solutions for isolated systems.  The time evolution solver was tested against \texttt{RAMSES} \citep{2002A&A...385..337T} and one of its MG versions \citep{2014A&A...562A..78L}.  The initial conditions were tested against the package COSMICS \citep{1995astro.ph..6070B} and the power spectrum calculator results were compared with Powmes \citep{2011ascl.soft10017C}.  Satisfactory results were found in all the cases.

The method that the code uses to run cosmological simulations is particle mesh with a uniform grid.  The interpolation of the density and forces to and from the grid are made with the usual CIC or TSC\footnote{I remind the reader that same scheme has to be used to interpolate both quantities to ensure momentum conservation.} interpolations \citep{1988csup.book.....H}. The particles are evolved using a leap-frog algorithm with variable time steps and the parallelisation is OpenMP.  The philosophy behind the code is simplicity.  It was designed to be a light weight N-body code where new algorithms (such as the one presented here) can be tested in a simple way, without complications that arise from the implementation of adaptive mesh refinement or MPI parallelisation.  Once a method has been shown to work with this simple code, it can be implemented in state-of-the-art codes.

In order to implement the shrinking domain framework, we need to be able to flag those cells that belong to the active part of the box (i.e. those that are inside the lightcone).  The position of the boundary of the lightcone when working with conformal time is given by
\be
r_{\mathrm{lc}}(t)  = c (t-t_0), 
\ee
where $t_0$ corresponds to the initial time of the simulation.

The code identifies the active cells simply by fixing the initial and final indexes of the loops in each direction.  Optimized versions of the code will deallocate the non-active cells.  However, in this particular implementation, the whole grid is kept in memory during the entire simulation (even if only a few cells are active at low redshift).  The same approach was applied to the particles (i.e. the code keeps all the particles in memory while the box shrinks and has an array of pointers that indicates the active particles).  The pointers are recalculated at the end of every time step, when the position of the boundary of the lightcone changes.

Three different versions of the code will be used in the following section to test the shrinking domain framework:
\bi
\item \texttt{Solve-std}:  uses a standard algorithm that solves Poisson's equation in the static limit in a periodic box.
\item \texttt{Solve-diffusion}: uses a standard algorithm that involves a diffusion equation for gravity in a periodic box.
\item \texttt{Solve-shrink}: uses the shrinking domain framework that involves a diffusion equation for gravity in a domain that shrinks with time.
\ei
The three codes use the same way of accessing cells in the grid and particles.  Also the memory allocation of these quantities is identical.  In this way, it is possible to isolate the effects that shrinking the box has on the CPU time without having different biases that may occur from cache misses that could exist in one implementation and not in others.  At the end of the day, this will be implemented in AMR codes, for which the dynamic allocation of independent cells and particles is already in place, so no overhead should exist in a state-of-the-art code that implements the shrinking domain framework.  The codes \texttt{Solve-diffusion} and \texttt{Solve-shrink} differ \textit{only} in the existence of a routine that eliminates particles and cells.

\begin{table}
\centering
\caption{Comoving box size and number of particles of the simulations used for testing.}
\label{table:simulations}
\begin{tabular}{cc}
Box (Mpc/$h$) & $N_{\mathrm{particles}}$ \\
\hline
512 & $512^3$ \\
2048 & $512^3$ \\
8192 & $512^3$ \\
8192 & $1024^3$ \\
\end{tabular}
\end{table}

\begin{figure*}
  \begin{center}
    \includegraphics[width=\textwidth]{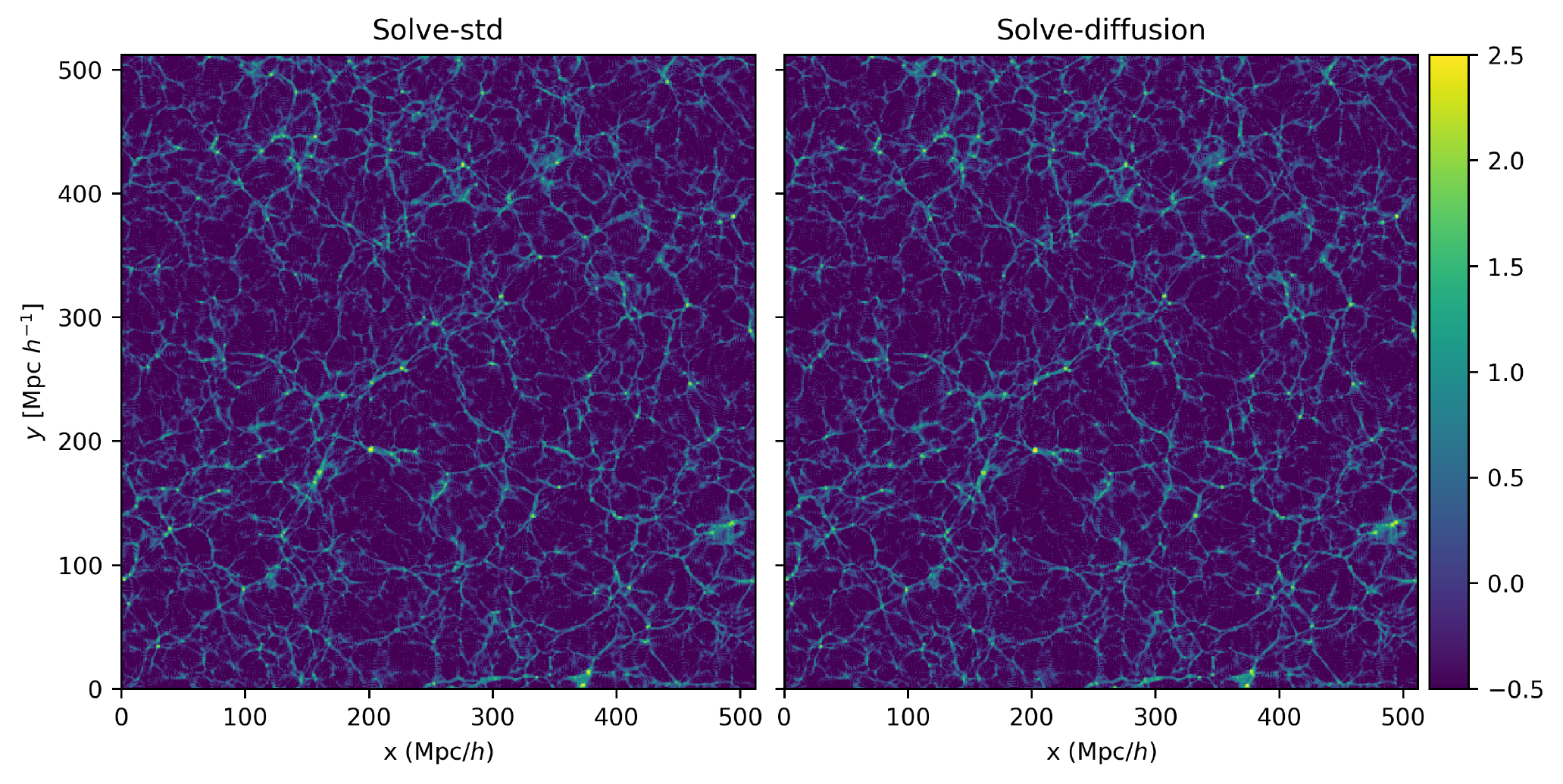}
    \caption{Distributions of log-density at redshift z = 0 in a plane that passes through the centre of the box that was run with $512^3$ particles and box size of 512 Mpc/$h$.  The left and right panels correspond to the results obtained with the codes \texttt{Solve-std} and \texttt{Solve-diffusion}.   A quantitative comparison of these two snapshots is shown in Figure \ref{fig:spectra_std_force}.}
    \label{fig:maps}
  \end{center}
\end{figure*}

\section{Tests}
\label{section:tests}

The aim is to present a new framework for cosmological simulations and test both its accuracy and its performance with respect to the standard method.  The test presented here consists of two stages.  First, we need to test that the non-static solver used for solving the diffusion equation for gravity is accurate enough and of comparable speed with respect to the usual Poisson solvers.  Only after doing this, we can activate the shrinking domain and compare both, accuracy and speed with respect to the diffusion solver.  So these two stages will done by comparing the codes \texttt{Solve-std} and \texttt{Solve-diffusion} first and then \texttt{Solve-diffusion} and \texttt{Solve-shrink}.

The tests are based on a set of 12 simulations (four different combinations of box size and number of particles run with the three different codes).  Details of these simulations are summarized in Table \ref{table:simulations}.  The comoving box sizes at the initial conditions range from 512 Mpc/$h$ to 8192 Mpc/$h$.  For each box size there is a simulation with $512^3$ particles.  Furthermore, the data set contains a simulation that was run with $1024^3$ particles (with the largest box size) which is intended to study the dependence of the results with the number of particles.  

The time integration variable is conformal time and 1000 steps uniform in $a$ were taken during the simulations.  This means that the time steps become smaller and smaller as time passes, which is required to properly calculate the internal dynamics of objects at late times.  The reason why I choose to have the same number of time steps for all the simulations (independently of the box size) is that in this way I can reduce the number of free parameters and isolate the effects on performance of the shrinking domain and thus, be able to make a clear comparison between the different methods.  When the time comes to include this method in state-of-the-art codes, the number of steps will be of the order of $10^3$ even for very large boxes \citep[e.g.][]{2008MNRAS.391..435F, 2012MNRAS.426.2046A, 2015MNRAS.448.2987F}.

The initial conditions were calculated at redshift $z=40$ with the initial conditions generator of the \texttt{Solve} package using the Zeldovich approximation.  The power spectrum used for this was obtained with the \texttt{COSMICS} package \citep{1995astro.ph..6070B}.  Note that the box sizes used for the simulations result in the boxes being well within the lightcone at the initial redshift.  So the simulations were run using periodic boundary conditions until the lightcone was small enough to reach the borders of the box.  Only when that happens does the code \texttt{Solve-shrink} start shrinking the box.

All the simulations were run with 60 cores of the same shared memory machine, so we can ensure that differences in CPU time arise only due to differences in the codes and not because of changing the architecture of the computer.

\subsection{Testing the diffusion solver:  \texttt{Solve-std} vs. \texttt{Solve-diffusion}}

This section presents the first part of the whole test, which consists of measuring changes in accuracy and performance that occur because of changing the gravitational solver. In the next section the effects of changing the box size on the fly will be measured.

\subsubsection{Accuracy of the diffusion solver}

Although the ADE solver used to integrate the diffusion equation for gravity is unconditionally stable, the special boundary conditions break the symmetry of the algorithm and thus impose a constraint on the number of time steps (or the speed of light).  While this constraint in not as stringent as the one provided by the CFL condition, it still requires the non-static simulations to have a larger number of time steps than expected.  The aim is not to find the best non-static solver but to test the shrinking domain framework.  So exact details of the solver used for the diffusion equation are not critical as long as it provides a solution that is good enough to test the shrinking box framework.  So a work around with this problem was implemented, which consists of reducing the speed of light for small boxes.  

Note that the reduction of the speed of light was made only when solving gravity.  The lightcone is still defined using the measured value and thus the timing results presented in following section were obtained in a realistic setup.  Future implementations of the non-static solver should certainly rely on implicit methods, which I leave for future work \citep[see for instance][for an example of a implicit solver with periodic boundary conditions]{2016JCAP...07..053A}.  See also \cite{2016PhRvD..93h3006H} for an example of the use of a hyperbolic solver to deal with gravity.

The CFL condition says that information, which travels at the speed of light, should not travel more than the size of a cell $h$ in time step $\Delta t$:
\be
c \Delta t < h.
\ee
In order to make a one to one comparison of the codes, I decided to keep the number of time steps fixed for all the simulations and change the value of the speed of light with the following phenomenological relation:
\be
c_{\mathrm{eff}} = 0.019 h c, 
\ee
where $c_{\mathrm{eff}}$ is the effective speed of sound used in the code and $h$ is the comoving size of cells in the grid.  The relation was obtained by running simulations with different values of $c_{\mathrm{eff}}$ and testing for the stability of the solutions.  The goodness of this criteria can only be assessed by comparing simulations run with the standard Poisson's method and the non-static equation (i.e. the codes \texttt{Solve-std} and \texttt{Solve-diffusion}).

\begin{figure}
  \begin{center}
    \includegraphics[width=.48\textwidth]{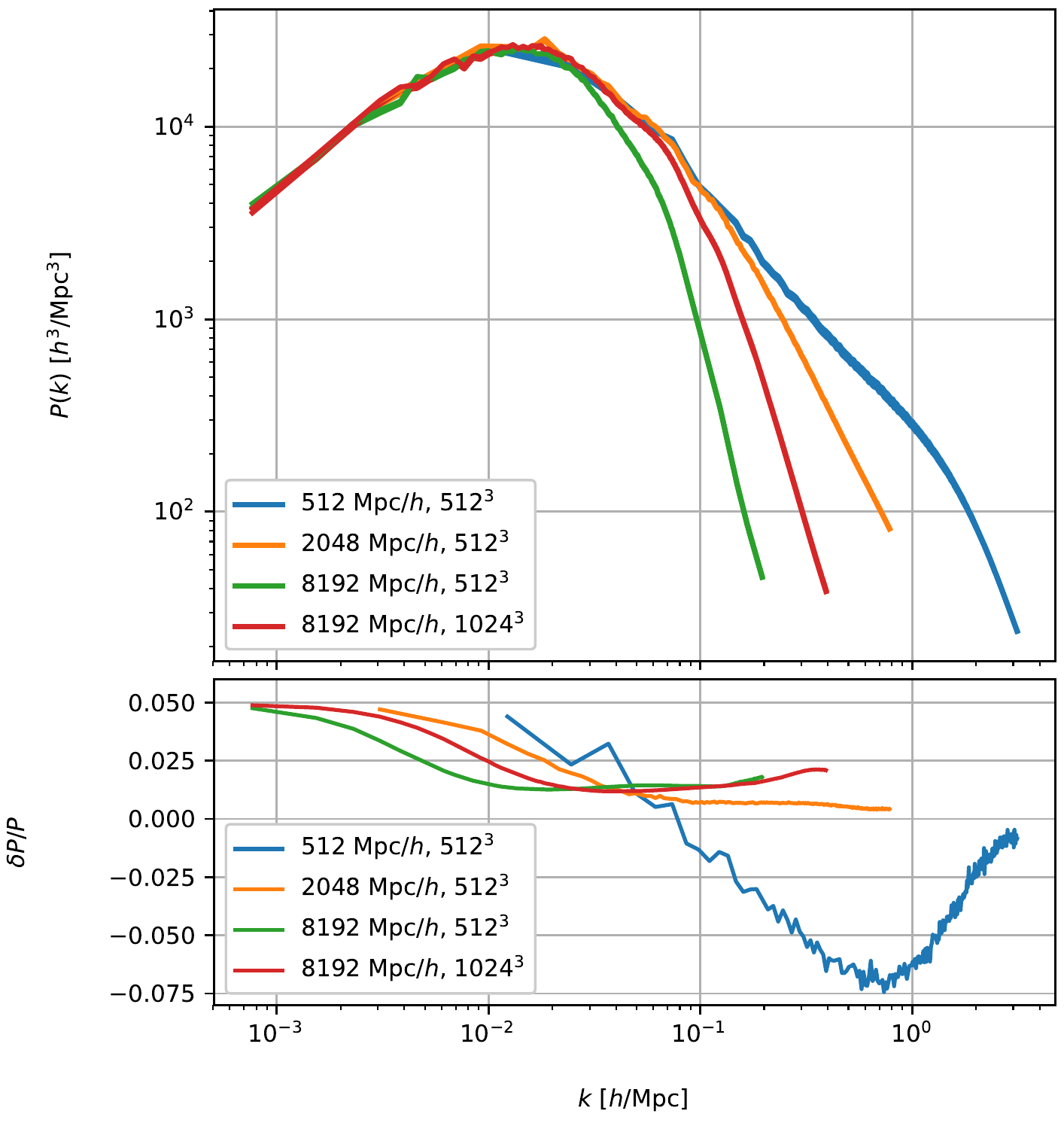}
    \caption{Comparison of spectra obtained with the standard method using a static Poisson's equation and the ADE method applied to a diffusion equation.  Top:  spectra of all the simulations.  Bottom:  relative difference between the two approaches.  The comoving size of the box is fixed in all these simulations.} 
    \label{fig:spectra_std_force}
  \end{center}
\end{figure}

The first thing that has to be checked when comparing cosmological codes is if the cosmological box actually looks like a cosmological box. This is important because even if we can prove that the power spectra obtained with different codes agree with each other, the phases could be wrong and give a weirdly shaped cosmic web.  As the box of these simulations does not shrinking, we can compare snapshots at redshift zero of the complete box, as is done when working with the usual approach.  Figure \ref{fig:maps} shows distribution of log-density for the higher resolution simulations (run with $512^3$ particles in a box of 512 Mpc/$h$) and for a slice that passes through the centre of the box.  Left is the result of the usual approach and right the one that was obtained with the non-static diffusion solver.  The distribution of halos, filaments and voids is very similar for both simulations, which shows that solving a diffusion equation instead of Poisson's equation can recover features in the cosmic web with great accuracy.

To give a quantitative description of the agreement between the results obtained with the static and non-static codes, we need to calculate a statistical quantity such as the power spectrum of density fluctuations.  The result of this comparison is shown in Figure \ref{fig:spectra_std_force}.  The upper panel shows the spectra measured from the eight simulations.  There are only minor differences between the two codes, so the plot appears to have only four curves, which correspond to the different box sizes and number of particles.  The high resolution box can recover the usual non-linear feature at $k\sim 1$.  In the rest of the boxes, the resolution is too low to recover any non-linear feature.  By comparing the two curves that correspond to the box of 8192 Mpc/$h$ (run with $512^3$ and $1024^3$ particles) it is possible to confirm that the lack of power at small scales is associated with a lack of resolution.  While the spectra of these very large boxes is unrealistic, running these boxes is important because effects related to the horizon when shrinking the box will only appear on them.

The lower panel of the same figure shows the relative difference between these spectra.  While differences are not zero, the deviations are at the few per cent level, which we will consider as good enough for the actual comparison we want to make in this paper, which is related to the time evolution of the box size.  We deal with this comparison in the following sections.

\subsubsection{Performance of the diffusion solver}

The codes \texttt{Solve-std} and \texttt{Solve-diffusion} were written such that they are as similar as possible to each other.  This will permit us to measure differences in running time that are strictly related to the methodology rather than technicalities of the implementation.  However, there are two main reasons why comparing wall times obtained with these two codes could be unfair:
\bi
\item The non-static solver is not fully parallelized.  The reason for this is that the ADE algorithm requires every cell to be able to access the updated values of some of its neighbours, so a simple OpenMP parallelization (such as the one that is implemented in \texttt{Solve}) is not possible.
\item While the two codes are part of the same package, the gravity solvers are completely independent and were optimized in different ways.  So differences may arise simply because the routines of one solver have more or less cache misses than the other.  This may vary when dealing with different codes and thus, these differences cannot be taken as a measurement of the performance of the methods.
\ei
Nevertheless, since our aim is to compare the relative performance (and accuracy) of the codes \texttt{Solve-diffusion} and \texttt{Solve-shrink} it is still important to have approximate measurements of how \texttt{Solve-std} and \texttt{Solve-diffusion} compare. Three runs with $512^3$ particles were made which each code (corresponding to three different spatial resolutions).  The total CPU time in hours taken by the three runs together is the following:
\begin{align*}
9.49~\mathrm{hours} & ~~~ \leftarrow ~~~ \mathrm{\texttt{Solve-std}} \\
10.82~\mathrm{hours} & ~~~ \leftarrow ~~~\mathrm{\texttt{Solve-diffusion}~(not~fully~paralellized)} \\
7.00~\mathrm{hours} & ~~~ \leftarrow ~~~\mathrm{\texttt{Solve-diffusion}~(fully~parallelized)} 
\end{align*}
As the code does not have a refinement structure, the real elapsed time per time step is the same throughout the simulations.  Furthermore, as the number of time steps was fixed to the same value for all the simulations, the times shown here are roughly three times the time taken by each simulation separately.  The third line above corresponds to the code \texttt{Solve-diffusion} when it is fully parallelized.  As the ordering of the operations is not the correct one in a simple OpenMP implementation, the solution departs slightly for the actual solution and thus, the fully parallelized implementation can not be used for science.  However, this version can give an estimation of how a properly parallelized MPI code will behave.  Note that the comparison of accuracy presented in previous section was not made with the fully parallelised version, but with the original one, in which the solution of the diffusion equation is not parallelised.  Thus, the solutions used for the test are the most accurate solutions the ADE method can give.

\begin{figure}
  \begin{center}
    \includegraphics[width=0.48\textwidth]{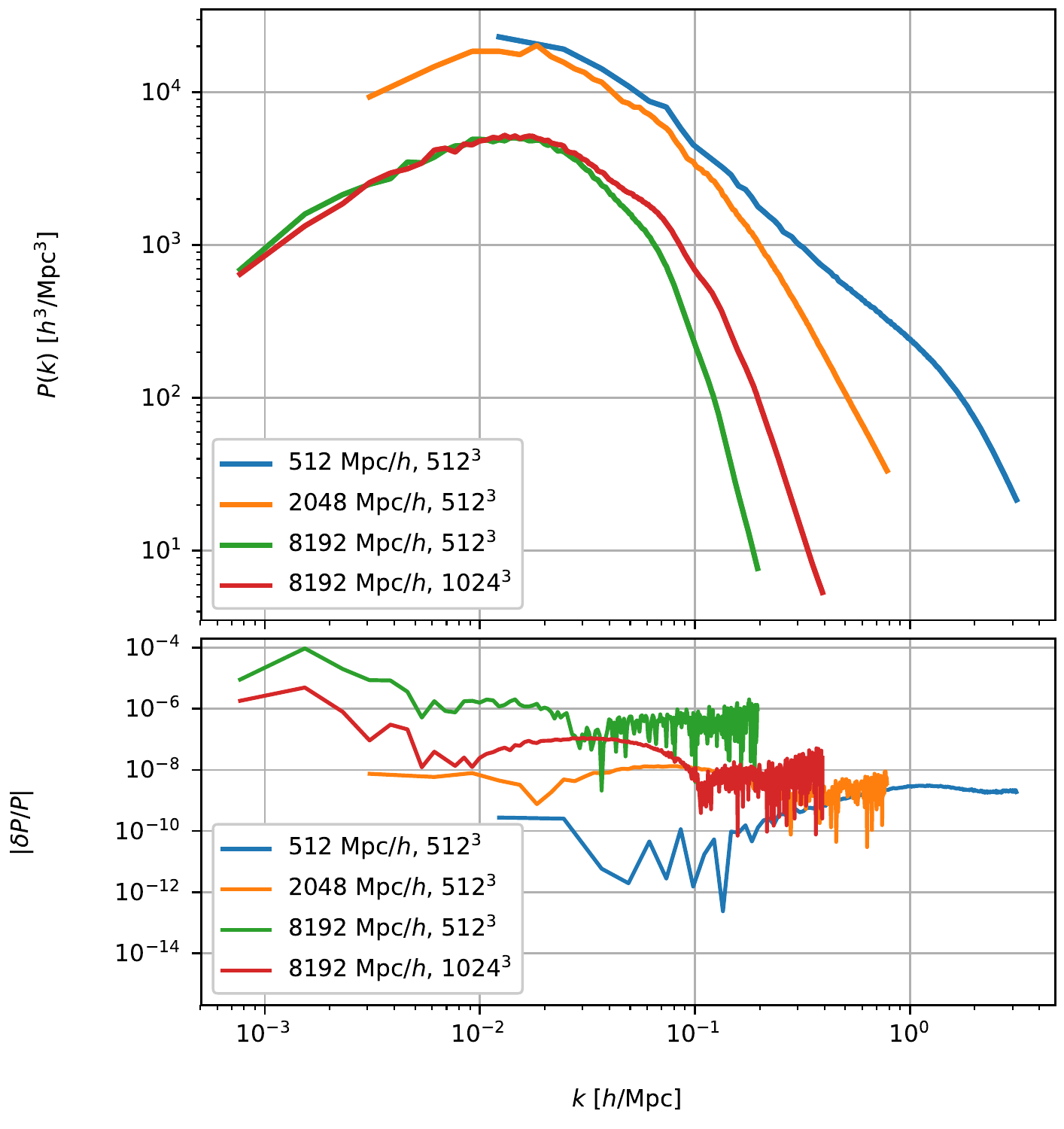}
    \caption{Comparison of spectra of lightcones obtained with the two versions of the diffusion code (i.e. with and without shrinking the simulation domain).  Top:  spectra of all the simulations.  There appear to be only four curves (instead of eight) because lines produced with different codes lie on top of each other.  For the larger boxes, the lightcone includes objects at high redshift, so their associated spectra have lower normalization that the ones of the small boxes.  Bottom:  relative difference between the two different approaches.}
    \label{fig:spectra_for_shr}
  \end{center}
\end{figure}

The final result of the comparison presented in this section is that the ADE method is competitive with the standard Poisson code and might even outperform it in terms of speed.  This allow us to go ahead and test what is the improvement in performance when shrinking the box size following the lightcone of the observer, which we do in the following sub-section.

\begin{figure}
  \begin{center}
    \includegraphics[width=0.48\textwidth]{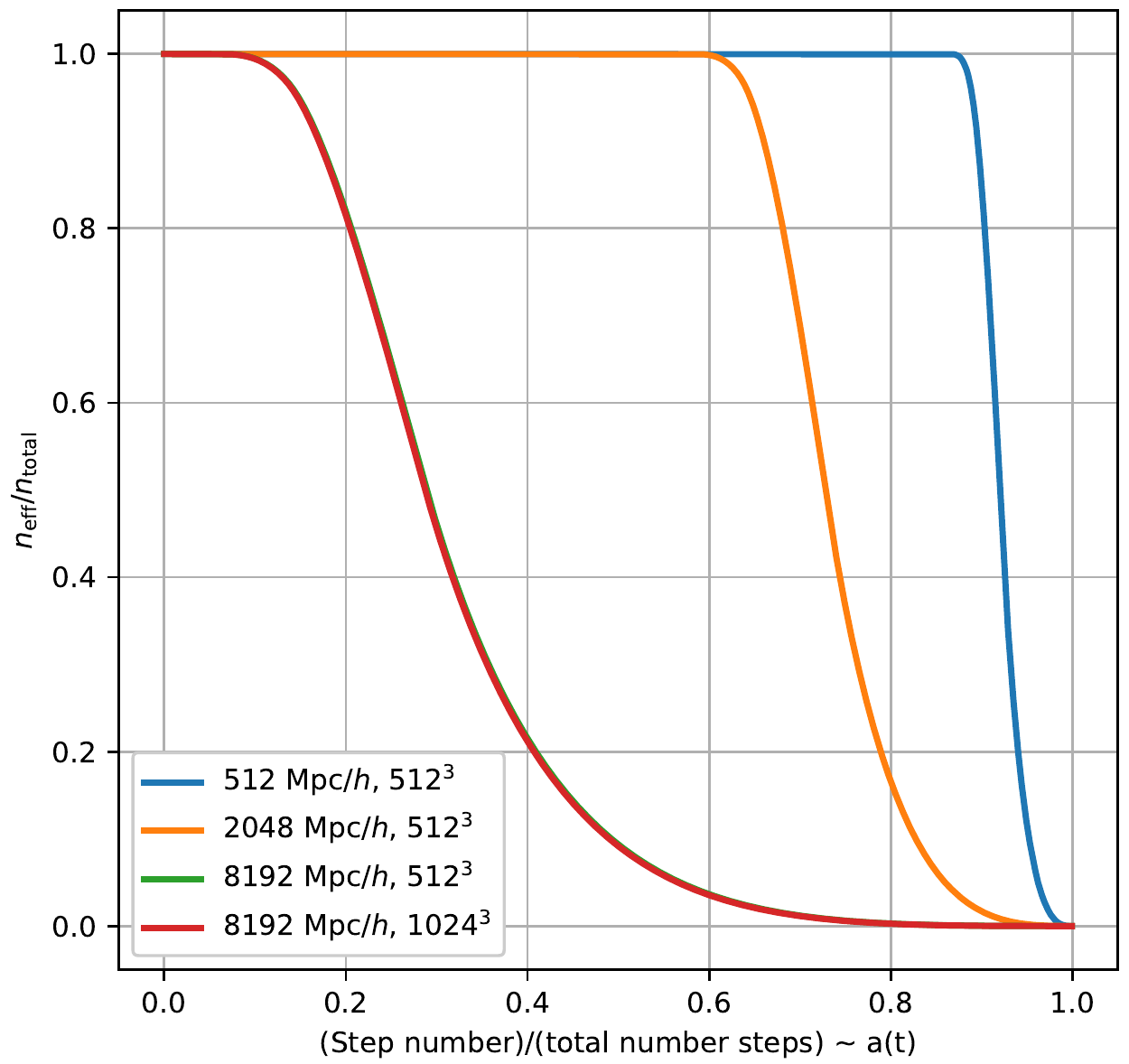}
    \caption{Number of particles integrated per time step in the simulations that shrink the box.  Different curves correspond to different box sizes and number of particles.  The curves that correspond to the two boxes with initial comoving box size of 8192 Mpc/$h$ lie in top of each other and thus, the plot contains only three curves for four simulations.}
    \label{fig:neff_of_nstep}
  \end{center}
\end{figure}

\subsection{Testing the shrinking domain framework:  \texttt{Solve-diffusion} vs. \texttt{Solve-shrink}}

This section contains the main result of the paper, which consists of measuring the performance of the shrinking domain framework.  Following the previous section, we check first that shrinking the box does not change the physics of the simulation in the observable region (i.e. that \texttt{Solve-shrink} can recover exactly the same power spectrum provided by \texttt{Solve-diffusion}).  Secondly, we measure the impact on performance of reducing the box size with time.

\begin{figure*}
  \begin{center}
    \includegraphics[width=\textwidth]{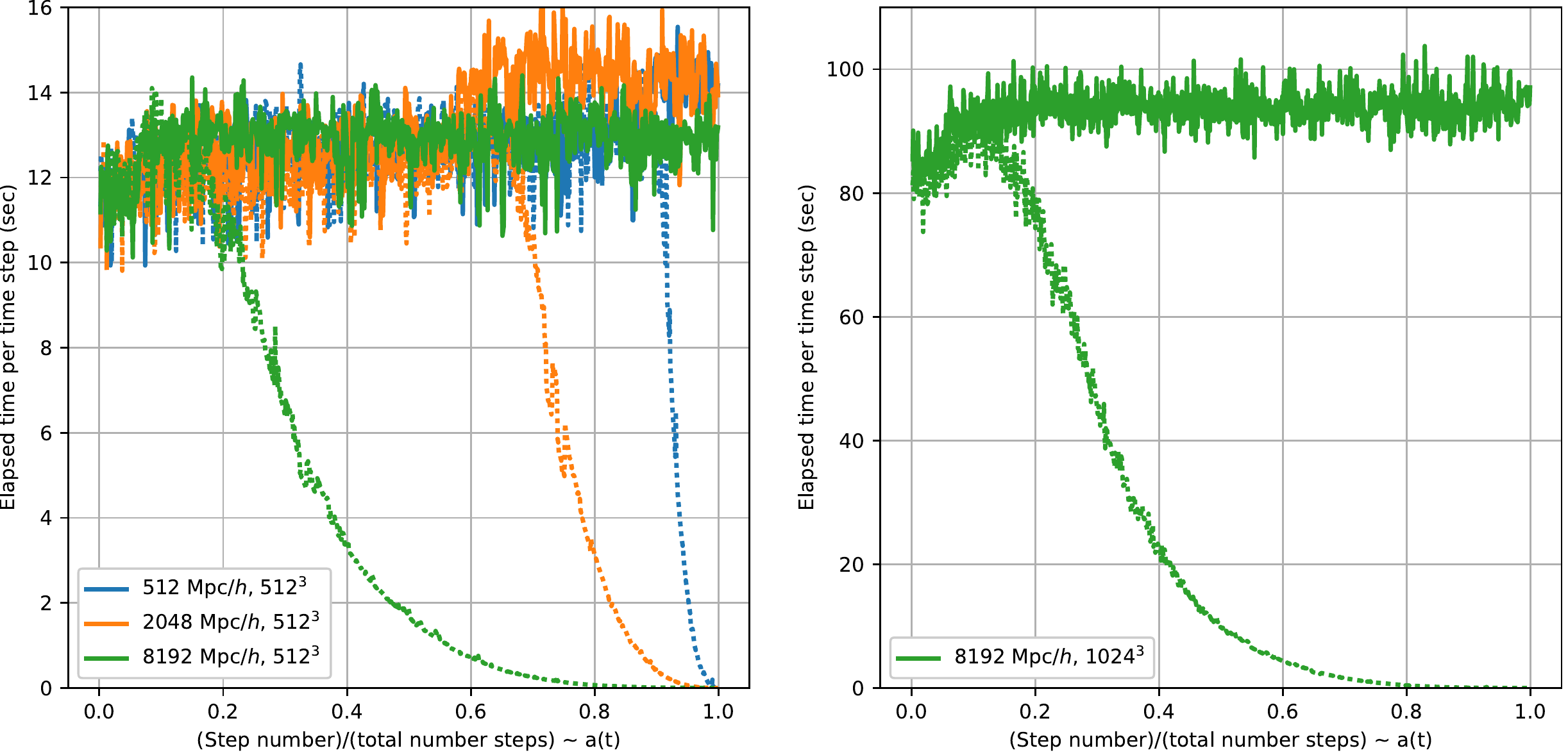}
    \caption{Elapsed time per time step for the simulations with $512^3$ (left) and $1024^3$ (right) particles.  The horizontal axis corresponds to the step number in terms of the total number of steps.    The continuous and dashed curves correspond to the simulations with fixed and shrinking domains.  The change in speed of the codes at the time when the output of the lightcone starts is related to the fact that the routine that takes care of it (which is not active before that moment) is not parallelized.} 
    \label{fig:elapsed_time}
  \end{center}
\end{figure*}

\subsubsection{Accuracy of the shrinking domain code}

When the routine that reduces the size of the box is active, the code cannot output snapshots anymore (simply because they are not calculated for the entire box).  The only output that we can compare is the lightcone.  For simplicity, I compare the 3D power spectra calculated in a cubical box of the original size which contains the whole lightcone.  Figure \ref{fig:spectra_for_shr} shows the result of this comparison.  The upper panel shows the 3D power spectra of all eight simulations (four different combinations of initial box size and number of particles for each of the two codes \texttt{Solve-diffusion} vs. \texttt{Solve-shrink}).  There appear to be four curves in the plot because simulations run with different codes give results that are almost exactly the same (see relative difference in the bottom panel of the same figure).

In the smaller box, the part of the code that shrinks the domain is activated only at very low redshift, and thus, the simulation is almost a standard simulation (but with a diffusion solver).  Because of this, the codes recover almost exactly the usual power spectrum with the usual non linear feature at $k\sim 1 ~h$/Mpc.  For the larger boxes, the light cone includes very high redshifts objects and thus contains regions in which the density perturbations are small.  The 3D power spectrum then measures a combination of high and low redshift physics, which results in a reduce normalization.  Independently of this, both codes give results that agree better that one part in $10000$.  Comparison of the runs made with 8192 Mpc/$h$ box size and $512^3$ and $1024^3$ particles confirms again that the lack of power at small scales in these simulations is related to lack of resolution.

\begin{figure*}
  \begin{center}
    \includegraphics[width=\textwidth]{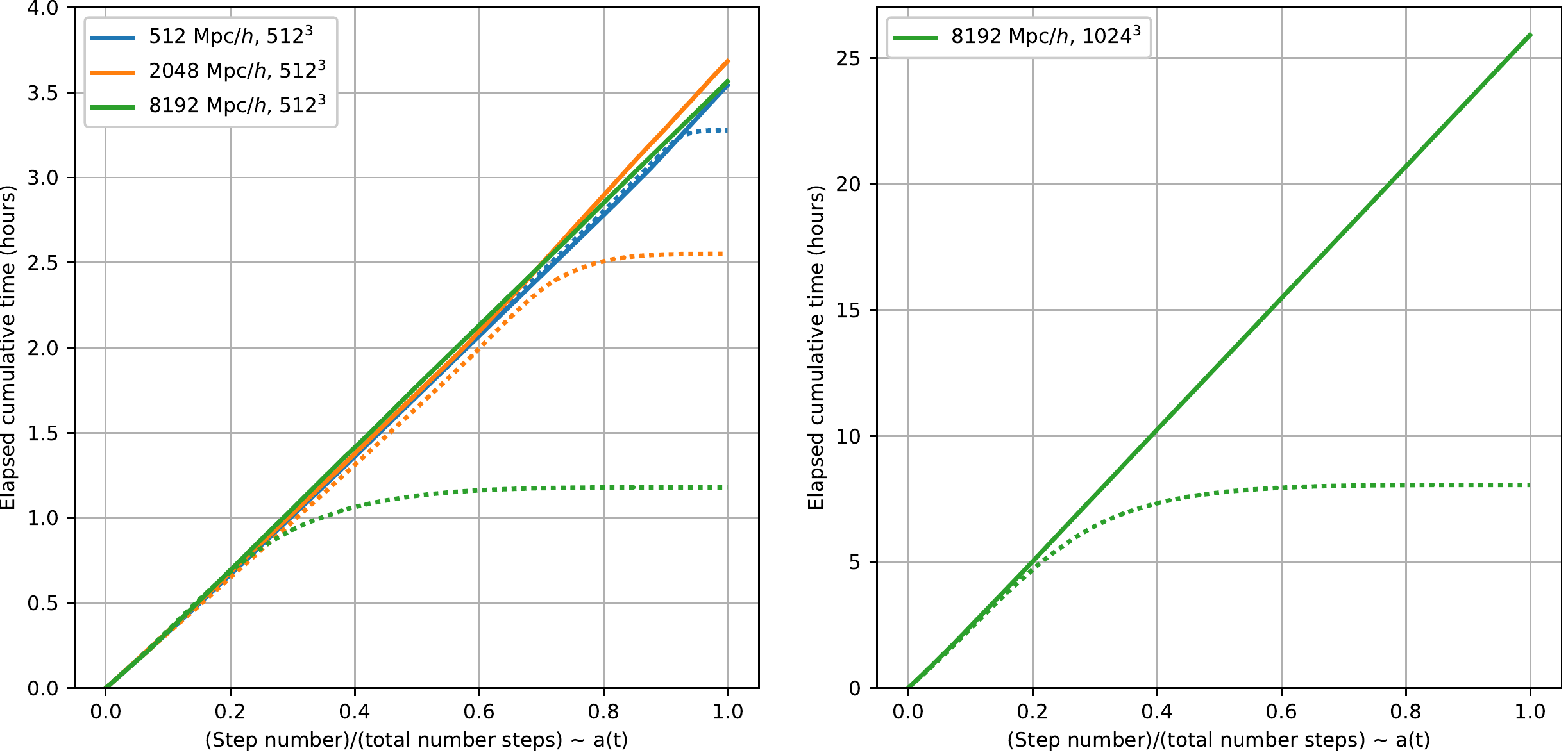}
    \caption{Cumulative elapsed time per time step (i.e. total time elapsed since the begining of the simulation as a function of time step) for the simulations with $512^3$ (left) and $1024^3$ (right) particles.  The continuous and dashed curves correspond to the simulations with fixed and shrinking domains.}
    \label{fig:cumulative_elapsed_time}
  \end{center}
\end{figure*}

\subsubsection{Performance of the shrinking domain code}

Now that we have demonstrated that shrinking the domain does not have any impact on the output of the simulations, we can finally measure the increase in the performance of the code, which is expected to occur because the smaller the box, the smaller the number of particles that have to be integrated.  Figure \ref{fig:neff_of_nstep} shows this quantity for the four different simulations that were run with \texttt{Solve-shrink}.  The curves that correspond to the boxes of 8192 Mpc/$h$ (with $512^3$ and $1024^3$ particles) lie on top of each other, and thus only three curves are visible in the plot.  The horizontal axis is the time step number normalized to the total number of time steps, which is 1000 for all the simulations.  As the time steps are regular in $a$, this quantity is very close to the expansion factor.  The vertical axis shows the fraction of the original number of particles that is integrated in every time step.

The high resolution run has a small box to start with and thus, the shrinking domain algorithm is activated only at late times.  This can be seen in the blue curve, which has a value equal to one during almost the whole simulation.  The opposite happens for the largest box, which starts reducing the box size almost at the beginning of the simulation.  After one quarter of the total number of time steps, the number of particles is reduced already to one half of the original.  When the simulation is half way to finishing, the number of particles is already one order of magnitude smaller than the original, which makes the simulation extremely fast with respect to what it would be if it had to integrate all the particles until redshift zero.

Almost the exact same behaviour is obtained when measuring the elapsed time associated to each time step, which is shown in seconds in Figure \ref{fig:elapsed_time}.  The continuous line corresponds to the elapsed time per time step of the simulation run with the code \texttt{Solve-diffusion}.  As the code is a plain particle mesh code (with no refinement structure), it is expected that the elapsed time is the same for all time steps.  The oscillations that appear in the plot correspond to the natural oscillations of a cluster whose clock varies because of external factors such as the physical temperature of the computer nodes.

The dotted lines correspond to the simulations with the shrinking domain activated.  Left and right panels correspond to simulations ran with $512^3$ and $1024^3$ particles.  Same behaviour found in Figure \ref{fig:neff_of_nstep} is found here:  The high resolution simulation (with smaller box size) shows a reduction of the elapsed time per time step only at the end of the simulation.  On the other hand, the very large box starts decreasing the elapsed time right at the beginning of the simulation.  After the simulation finished integrating one quarter of the total number of time steps, the elapsed time per time step has reduced by half and when the simulation integrated half of the time steps, the elapsed time reduced by an order of magnitude with respect to the original.

The impact of all this on the total CPU time required per simulation is shown in Figure \ref{fig:cumulative_elapsed_time}, which has the same horizontal axis as Figure \ref{fig:elapsed_time}, but shows the integrated elapsed time.  The straight lines correspond to the simulations run with \texttt{Solve-diffusion} and the lines that become horizontal where obtained with \texttt{Solve-shrink}.  The very large box, has an improvement of the total speed of a factor of about three by the end of the simulation.  Comparison of the left and right panel shows that this result is independent of the number of particles.

\begin{table}
\centering
\caption{Ratio between the total CPU time of the simulations ran with the codes \texttt{Solve-diffusion} and \texttt{Solve-shrink}, which shows that total improvement in speed when reducing the size of the domain on the fly.}
\label{table:speed}
\begin{tabular}{ccc}
Box (Mpc/$h$) & $N_{\mathrm{particles}}$ & Speed up \\
\hline
512 & $512^3$ & 1.08\\
2048 & $512^3$ & 1.44\\
8192 & $512^3$ & 3.02\\
8192 & $1024^3$ & 3.21 \\
\end{tabular}
\end{table}

The final speed up obtained by shrinking the domain is show in Table \ref{table:speed} for all the simulations.  It is important to remind the reader that these simulations were run with a particle mesh code with a uniform mesh and uniform time steps in $a$ and so, these numbers should to be taken only as indicative lower bounds.  In a state-of-the-art code, the situation is different.  The code will increase time resolution locally at late times when structures start to depart from linearity and thus, the elapsed time per time step will not be uniform (as Figure \ref{fig:elapsed_time} shows), but will increase.  These are the times when the number of particles is low, so the improvement in speed is expected to be much larger than found here.

\section{Conclusions}
\label{section:conclusions}

Large N-body simulations are necessary, but very expensive.  In an attempt at reducing the costs of the simulations in terms of CPU time and memory requirements, I propose a new framework for running cosmological simulations, which consists of changing the boundary conditions from periodic to open and using a time dependent comoving box size which follows the lightcone of an observer at redshift zero.

Thanks to the new geometry of the box, standard Poisson solvers with periodic boundary conditions are not appropriate anymore to calculate gravitational forces.  Instead, I propose to use the solution of a generalized Poisson's equation, which has the form of a diffusion equation.

To test that the new proposed framework gives reliable results in a shorter amount of CPU time, I implemented a solver for the generalized Poisson's equation and the shrinking domain framework in the N-body code \texttt{Solve} and ran simulations to compare the outcome of the standard and new framework.  Two comparisons were made:  between a standard Poisson code and a code that includes the diffusion equation and between this second code and the same code with the shrinking domain framework implemented.  The three codes were compared using the final power spectrum of density perturbations.  The solver used to integrate the diffusion equation gives a solution whose accuracy is better than 7.5 \% in all the scales studied.  On the other hand, the relative difference between this code and the one that shrinks the box is below $10^{-4}$ showing that shrinking the box size while the simulation runs, has no effect at all in the final outcome of the simulation and thus, it is a safe technique.  Note that the error of 7.5\% found when comparing the solver with the standard solver can be improved by changing the diffusion solver, but the difference of $10^{-4}$ found when implementing the shrinking domain is ready for science.

The CPU times required by the standard Poisson solver and the diffusion solver are similar and depend on details of the implementation and the parallelization strategies used.  However, the CPU time required by the code that includes the shrinking domain framework can be more than three times shorter for the larger boxes studied.  The code used for the test does not include adaptive mesh refinements, which are the main reponsible for the slow down of the simulations at low redshift.  Thus, these estimations are a conservative lower boundary for the speed up that will be obtained when using a state-of-the-art AMR code.  As the simulation uses an exact algorithm within the simulated region, the new proposed framework has the potential of giving a much higher speed up which does not come with the cost in accuracy that is associated to approximate methods.


\section*{Acknowledgements}

Thanks to Jose Beltr{\'a}n, David Mota, Carlton Baugh, Shaun Cole and Matthieu Schaller for discussions and carefully reading the manuscript.  Special thanks to Lydia Heck for support with computer issues.  This work used the DiRAC Data Centric system at Durham University.  This equipment was funded by BIS National E-infrastructure capital grant ST/K00042X/1, STFC capital grants ST/H008519/1 and ST/K00087X/1, STFC DiRAC Operations grant ST/K003267/1 and Durham University. DiRAC is part of the National E-Infrastructure.  I acknowledge support from STFC consolidated grant ST/L00075X/1 \& ST/P000541/1.

\bibliographystyle{mnras}
\bibliography{references}

\appendix
\section{Derivation of a generalized Poisson's equation}
\label{appendix:equations}

We are interested in finding a generalization of Poisson's equation, which can take into account the fact that information travels at finite speed through the box.  The only way to do this in a self consistent way is to included higher orders of an expansion of Einstein's equations (EE).  For our purposes, it will be enough to keep neglecting vector and tensor perturbations and work with the following metric in comoving coordinates:
\be
ds^2 = a^2\left[-\left(1+2\Phi\right)dt^2 + \left(1-2\Phi\right)\left(dx^2 + dy^2 + dz^2\right)\right], 
\ee
where $\Phi$ is a scalar perturbation over the Friedman background, $t$ is conformal time and $a$ is the expansion factor.  Poisson' equation will come out of the $00$ Einstein's equation, so we look only at these components of the tensors.  The $00$ component of the Einstein tensor is:
\begin{align}
G_{00} = & -\frac{6H \dot{\Phi}}{1-2\Phi} + \\ 
& 3 H^2 + \frac{2}{a^2(1-2\Phi)^2}\left[ (1+2\Phi)\nabla^2\Phi + 3 \dot{\Phi}^2  \right] + \\
& \frac{6(1+2\Phi)}{(1-2\Phi)^3} |\nabla\Phi|^2
\end{align}
where dots are derivatives with respect to conformal time, $H=\dot{a}/a$ and the symbols $\nabla^2$ and $\nabla$ are 3D gradient and Laplacian.   The same component of the energy-momentum tensor is:
\be
T_{00} = a^2 (\rho_0+\delta\rho)(1+2\Phi), 
\ee
where $\rho_0$ and $\delta\rho$ are background density and its perturbation respectively.  By plug in this in Einstein's equation
\be
G_{ab} + g_{ab}\Lambda = T_{ab}
\ee
and neglecting second order in the metric perturbations (not in the density) and keeping the time derivatives, we get:
\be
-3H\dot{\Phi} + \nabla^2\Phi - 3H^2\Phi = 4\pi G a^2\delta\rho.
\ee
For simplicity, we neglect also the term $3 H^2 \Phi$.  After substituting the factor $4\pi G$ by background quantities, we finally obtain the generalization we were looking for (i.e. Eq. \ref{diffusion}), which is the equation that I included in the N-body code.


\bsp    
\label{lastpage}

\end{document}